\documentclass{ws-procs975x65}

\begin{document}

\def\hm{\ \rm {\it h}^{-1} Mpc}

\def\ltsim{\, \lower2truept\hbox{${<
      \atop\hbox{\raise4truept\hbox{$\sim$}}}$}\,}

\def\gtsim{\, \lower2truept\hbox{${>
      \atop\hbox{\raise4truept\hbox{$\sim$}}}$}\,}

\def\omde{\Omega_{\rm DE}}
\def\cs2{c_s^2}

\title{DARK ENERGY CONSTRAINTS FROM NEEDLETS ANALYSIS OF WMAP3 AND NVSS DATA}

\author{Davide Pietrobon}
\address{Dipartimento di Fisica, Universit\`a di Roma ``Tor Vergata'',\\
V. della Ricerca Scientifica 1, I-00133 Roma, Italy\\
\email{davide.pietrobon@roma2.infn.it} }

\author{Amedeo Balbi}
\address{Dipartimento di Fisica, Universit\`a di Roma ``Tor Vergata''
and\\
INFN Sezione di Roma ``Tor Vergata'', V. della Ricerca Scientifica 1, I-00133 Roma, Italy\\
\email{amedeo.balbi@roma2.infn.it} }

\author{Domenico Marinucci}
\address{Dipartimento di Matematica, Universit\`a di Roma ``Tor Vergata'',\\
V. della Ricerca Scientifica 1, I-00133 Roma, Italy\\
\email{marinucc@mat.uniroma2.it} }


\begin{abstract}
We cross-correlate the new 3 year Wilkinson Microwave Anistropy Probe (WMAP3)  cosmic microwave background (CMB) data with the NRAO VLA Sky Survey (NVSS) radio galaxy data, and find further evidence of late integrated Sachs-Wolfe (ISW) effect taking place at late times in cosmic history. Our detection makes use of a novel statistical method based on a new construction of spherical wavelets, called needlets. The null hypothesis (no ISW) is excluded at more than 99.7\% confidence. When we compare the measured cross-correlation with the theoretical predictions of standard, flat cosmological models with a generalized dark energy component parameterized by its density, $\omde$, equation of state $w$ and speed of sound $\cs2$, we find $0.3\leq\omde\leq0.8$ at 95\% c.l., independently of $\cs2$ and $w$. If dark energy is assumed to be a cosmological constant ($w=-1$), the bound on density shrinks to $0.41\leq\omde\leq 0.79$. Models without dark energy are excluded at more than $4\sigma$. The bounds on $w$ depend rather strongly on the assumed value of $\cs2$.
\end{abstract}

\bodymatter

\section{Addressing the problem}
\label{intro}

The most outstanding problem in modern cosmology is understanding the mechanism that led to a recent epoch of accelerated expansion of the universe. The evidence that we live in an accelerating universe is now compelling. The most recent cosmological data sets, luminosity distance measured from type Ia supernovae \cite{Riessetal.2004}, the amount of clustered matter in the universe, detected from redshift surveys, clusters of galaxies, etc. \cite{Springeletal.2006}, and the pattern of anisotropy in the CMB radiation\cite{Spergeletal.2006} have shown that the total density of the universe is very close to its critical value, suggesting a flat universe geometry. Matter, characterized by ordinary actractive gravity, covers only $\sim 1/3$ of the total density, while the remaining $\sim 2/3$, responsable for the acceleration epoch, show a repulsive gravity. The precise nature of this cosmological term, however, remains mysterious. The favoured working hypothesis is to consider a dynamical, almost homogeneous component (termed {\em dark energy}) \cite{Caldwelletal.1998}.

One key indication of an accelerated phase in cosmic history is the signature from the integrated Sachs-Wolfe (ISW) effect \cite{Sachs&Wolfe1967} in the CMB angular power spectrum. A detection of a ISW signal in a flat universe is, in itself, a direct evidence of dark energy. Furthermore, the details of the ISW contribution depend on the physics of dark energy, in particular on its clustering properties, and are therefore a powerful tool to better understand its nature. Unfortunately, because of the cosmic variance, making the extraction of the ISW signal is a difficult task. A useful way to separate the ISW contribution from the total signal is to cross-correlate the CMB anisotropy pattern with tracers of the large scale structure (LSS) in the local universe.

In our analysis\cite{Pietrobonetal.2006} we improve previous studies in two directions. On one side we applied a new type of spherical wavelets, the so-called {\em needlets} \cite{Baldietal.2006b}, to extract cross-correlation signal beetwen WMAP3 CMB sky maps with the radio galaxy NVSS catalogue. Needlets have the great advantage to show a double localization: both in multipole space and in pixel space due to the analitic properties of the shape of the window function. This makes our analysis almost insensitive to the cuts present in the maps, increasing our ability in the detection of ISW effect. On the other side we treated dark energy as a generic fluid characterized by three physical parameters: its overall density $\omde$,  its equation of state $w=p/\rho$, and the sound speed $c_s^2= \delta p/\delta \rho$. This parameterization has the advantage of being model independent, allowing one to encompass a rather broad set of fundamental models, and of giving a more realistic description of the dark energy fluid.

The first step in our work was to build the two maps at the proper resolution, useful for our purpose, paying particular attention to remove noise and systematic errors; secondly we extract the cross-correlation coefficients from maps. To make a comparison with the theoretical prevision, we numerically integrated the Boltzmann equation for photon brightness coupled to the other relevant equations, suitably modifing the CMBFast code.

\section{Results}
\label{results}

In the left hand panel of the figure, we show the cross-correlation signal in wavelet space extracted from the WMAP3 and NVSS data, superimposed to the predicted cross-correlation for some dark energy models. The excess signal peaks at value $7<j<11$, corresponding to angular scales between $2^\circ$ and  $10^\circ$, as expected from theoretical studies. We checked that the observed signal was not produced by casual alignment of sources in the NVSS catalogue with the CMB pattern at decoupling ($z\simeq1100$) and the errors, calculated through the Monte Carlo procedure, are close to the analytical estimates, that confirm the consistence of our approach.

\begin{figure}
\centering
\includegraphics[width=0.45\columnwidth]{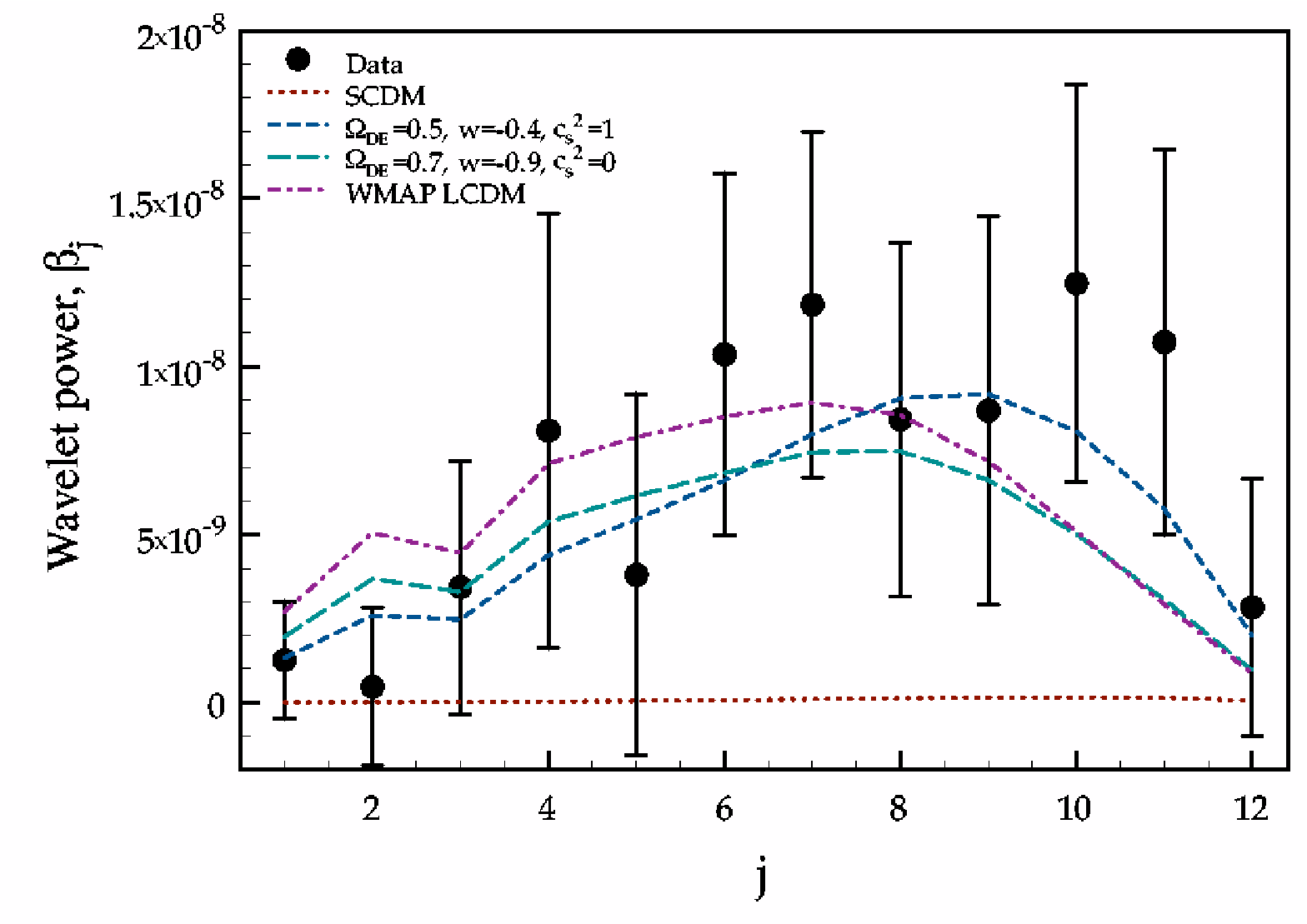}
\includegraphics[angle=90,width=0.45\columnwidth]{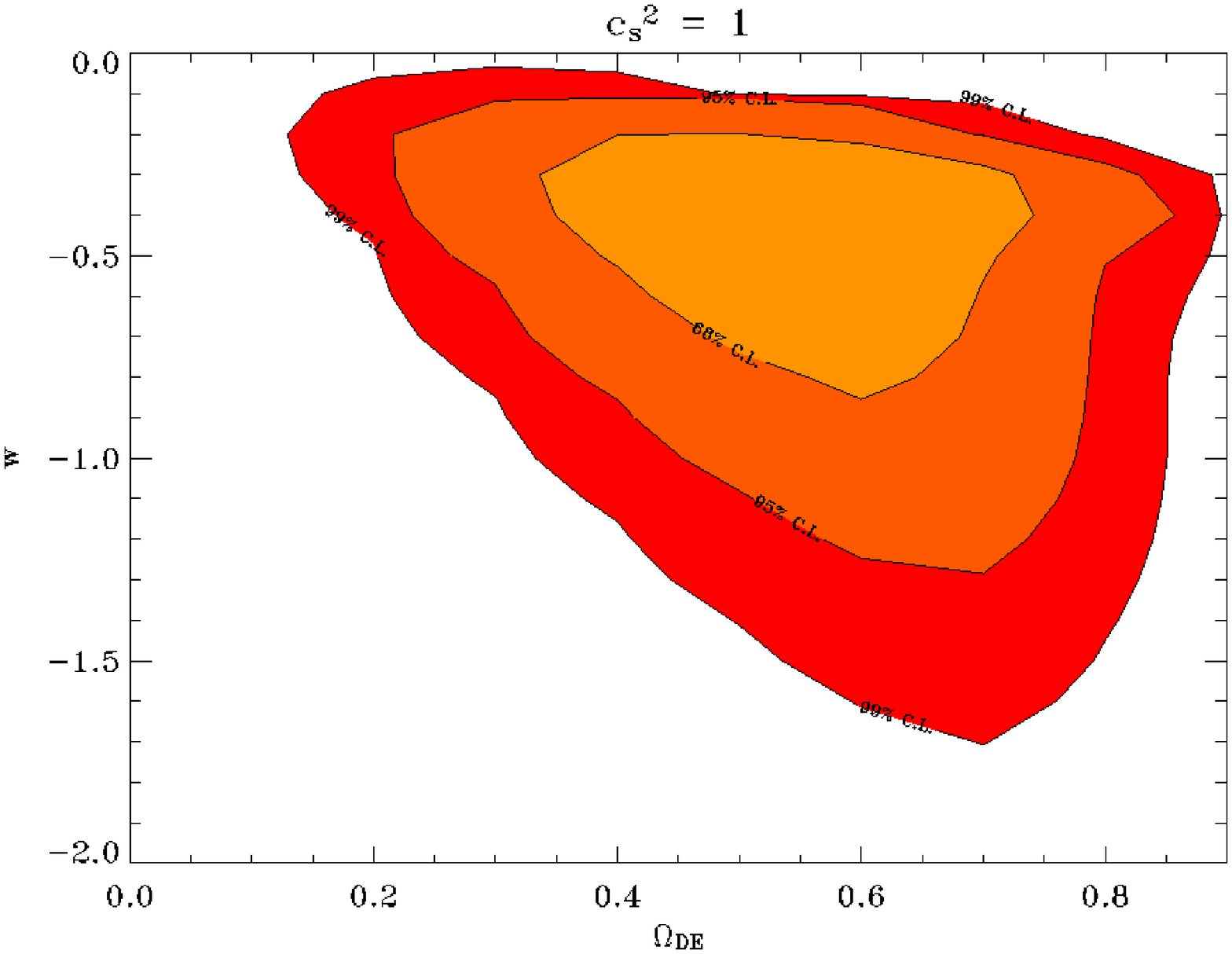}
\caption{Cross-correlation signal in needlet space extract from WMAP3 CMB data and NVSS radio galaxy survey (left hand panel). Note that a simple Cold Dark Matter model does not reproduce the signal (dotted line). in the right hand panel, constraints at 68\%, 95\% and 99\% confidence level in the $\omde$---$w$ plane for scalar field model (speed of sound $\cs2=1$). \label{fig:contours}}
\end{figure}

The first conclusion we can draw from our analysis is that the evidence for non zero dark energy density is rather robust: we find $0.32\leq\omde\leq 0.78$ at 95\% confidence level. A null value of $\omde$ is excluded at more than $4\sigma$, independently of $\cs2$. When we model the dark energy as a cosmological constant (i.e. we assume the value $w=-1$ for its equation of state), the bounds on its density shrinks to $0.41\leq\omde\leq 0.79$ at 95\% confidence level.

Quite interestingly, we find that although the case for a non zero dark energy contribution to the total density is compelling, the constraints on $w$ do depend on the assumed clustering properties of the dark energy component, namely its sound speed $\cs2$. Phantom models, and also the ordinary cosmological constant case $w=-1$, perform worse when a quintessence behaviour $\cs2=1$ is assumed. However, we emphasize that, for values of $\omde\sim 0.7$, models with $w=-1$ are a good fit to the data, as it is evident from the right hand panel in figure.

Clearly, the observation of ISW is proving quite promising as a tool to answer the questions arising from the mysterious nature of dark energy. While the CMB data have reached a great degree of accuracy on the angular scales that are more relevant for the detection of ISW, deeper redshift surveys and better catalogues can, in the future, improve the tracing of the local matter distribution, thus allowing to reduce the errors on the cross-correlation determination.

\vfill

\end{document}